\documentclass[12pt,preprint]{aastex}
\usepackage{graphicx,epsfig,epsf,amssymb}
\def \<{\langle}
\def \>{\rangle}

\newcommand{\degree}{^\circ}

\begin{document}

\title{New evidence for lack of CMB power on large scales}

\author{Hao Liu\altaffilmark{1} and Ti-Pei Li\altaffilmark{1,2,3}}
 \altaffiltext{1}{Key Lab. of Particle Astrophys., Inst. of High Energy Phys.,
Chinese Academy of Sciences, Beijing}
 \altaffiltext{2}{Department of Physics \& Center for Astrophysics,
Tsinghua University, Beijing, China}
\altaffiltext{3}{Department of Engineering Physics \& Center for Astrophysics,
Tsinghua University, Beijing, China}

\begin{abstract}
A digitalized temperature map is recovered from the first light sky
survey image published by the Planck team, from which an angular
power spectrum is derived. The amplitudes of the low multipoles 
measured from the preliminary Planck power spectrum are
significantly lower than that reported by the WMAP team. Possible systematical effects are far 
from enough to explain the observed low-$l$ differences.
\end{abstract}

\section{Introduction}
The first light survey image covering
about 10\% of the sky has been published by the Planck team$^{[1]}$. 
Recently, through a visual comparison
of this first light sky survey with the WMAP and COBE results
by using greyscale images, Cover$^{[2]}$  found that there are
substantial differences at large scale (low-$l$) between the WMAP
and Planck sky maps. In this work, we extract the temperature
intensity information from the published Planck image and located
their corresponding pixels on the sky to produce a sky map. 
We then derive a CMB power spectrum from the map and find 
 the amplitudes of the low multipoles 
measured from the spectrum being
significantly lower than WMAP release,
which is consistent with what is seen by Cover$^{[2]}$
and also consistent with our findings from an improved spectrum by
using the WMAP data and our own software$^{[3]}$.

\section{Producing temperature map}
\label{sec:From image to CMB temperature map}
\subsection{From image color to temperature bin number}
\label{sub:From image color to temperature bin number}

\begin{figure}[t]
\begin{center}
\includegraphics[width=0.6\textwidth]{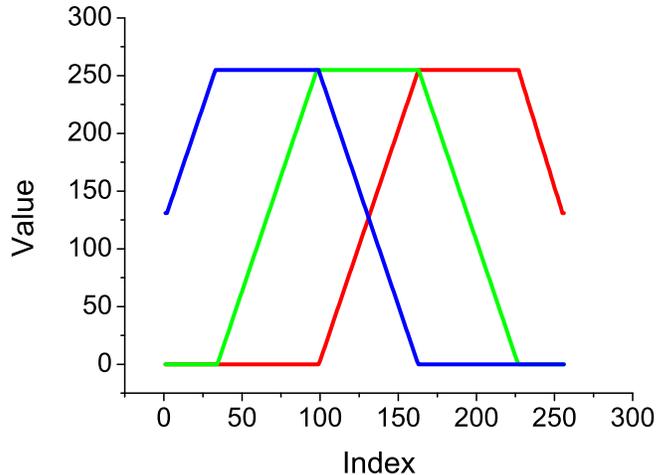}
\end{center}
\vspace{-4mm} \caption{The blue-to-red color table.
The horizontal axis is temperature bin number ($0-255$) between the maximum
and minimum temperatures, and the three curves in red, green and blue
give corresponding color values of each bin.}
\label{fig:color table}
\end{figure}

The published first light survey image$^{[1]}$ is plotted with a blue-to-red color table
on which lower values are presented with blue, median values with green,
and higher values with red (see Fig.~\ref{fig:color table}).
Therefore, we can obtain the temperature intensities by reversing this relationship.
Since the image is given in JPEG format, a lossy compression by which the color might
be slightly twisted, the safest way is to compute the rms difference of each pixel's
color to all 256 standard colors of 256 bins, and take the bin with the minimum rms
as the temperature intensity.

\subsection{From image to sky map}
\label{sub:From image to sky map}

The published image is plotted in Mollweide projection
in resolution $2048\times 1024$. We use the center of this image as the origin,
find corresponding HEALPix$^{[4]}$ pixel number of each image pixel,
and then set its value to the recovered temperature bin number ($0-255$).
Since the image contains no more than 2 millions available pixels,
which is much less than the effective pixel number of HEALPix 
resolution $N_{side}=512$,
the HEALPix resolution is finally set to $N_{side}=256$.

\begin{figure}[t]
\begin{center}
\includegraphics[width=0.6\textwidth]{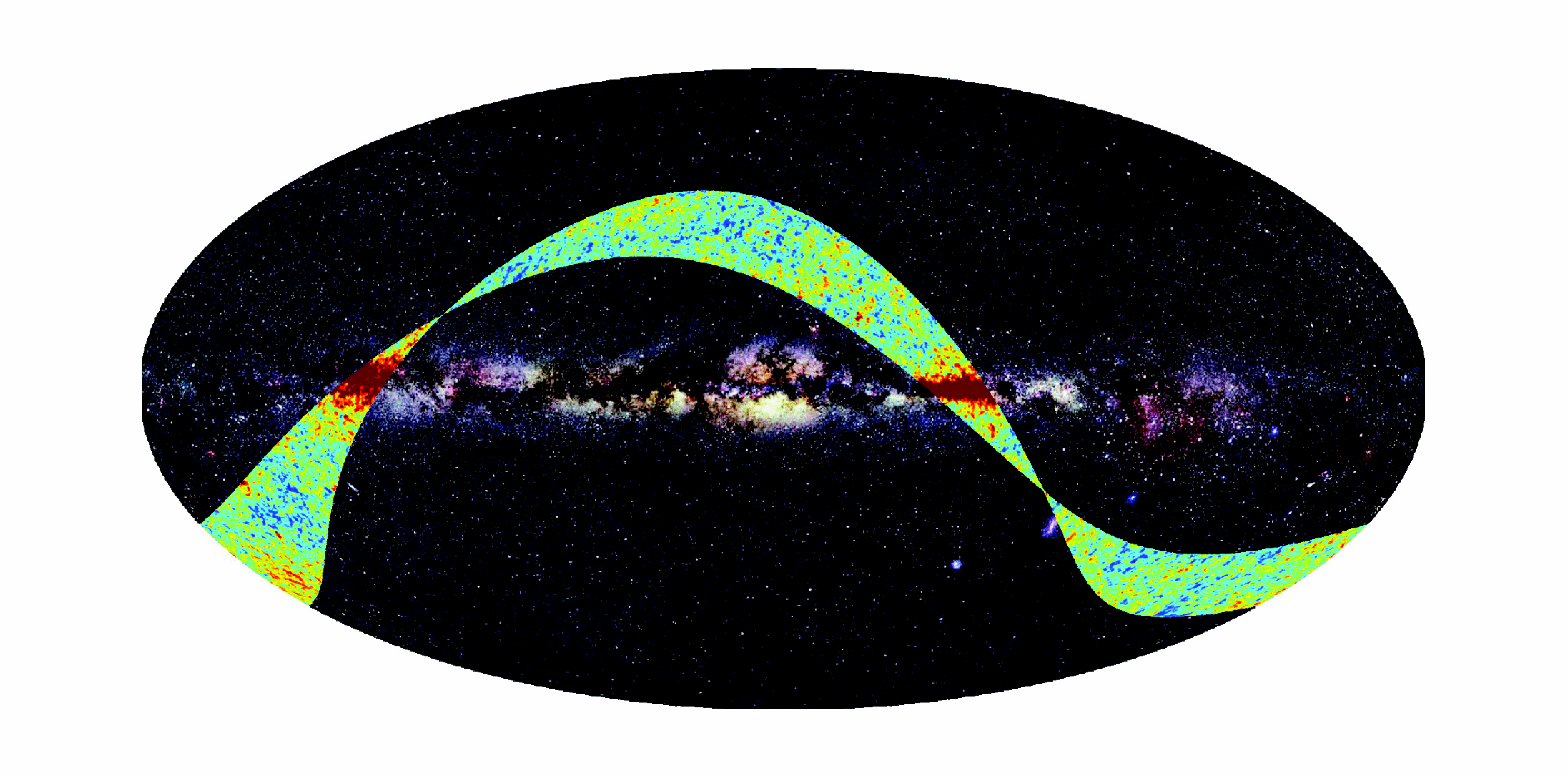}
\includegraphics[width=0.31\textwidth,angle=90]{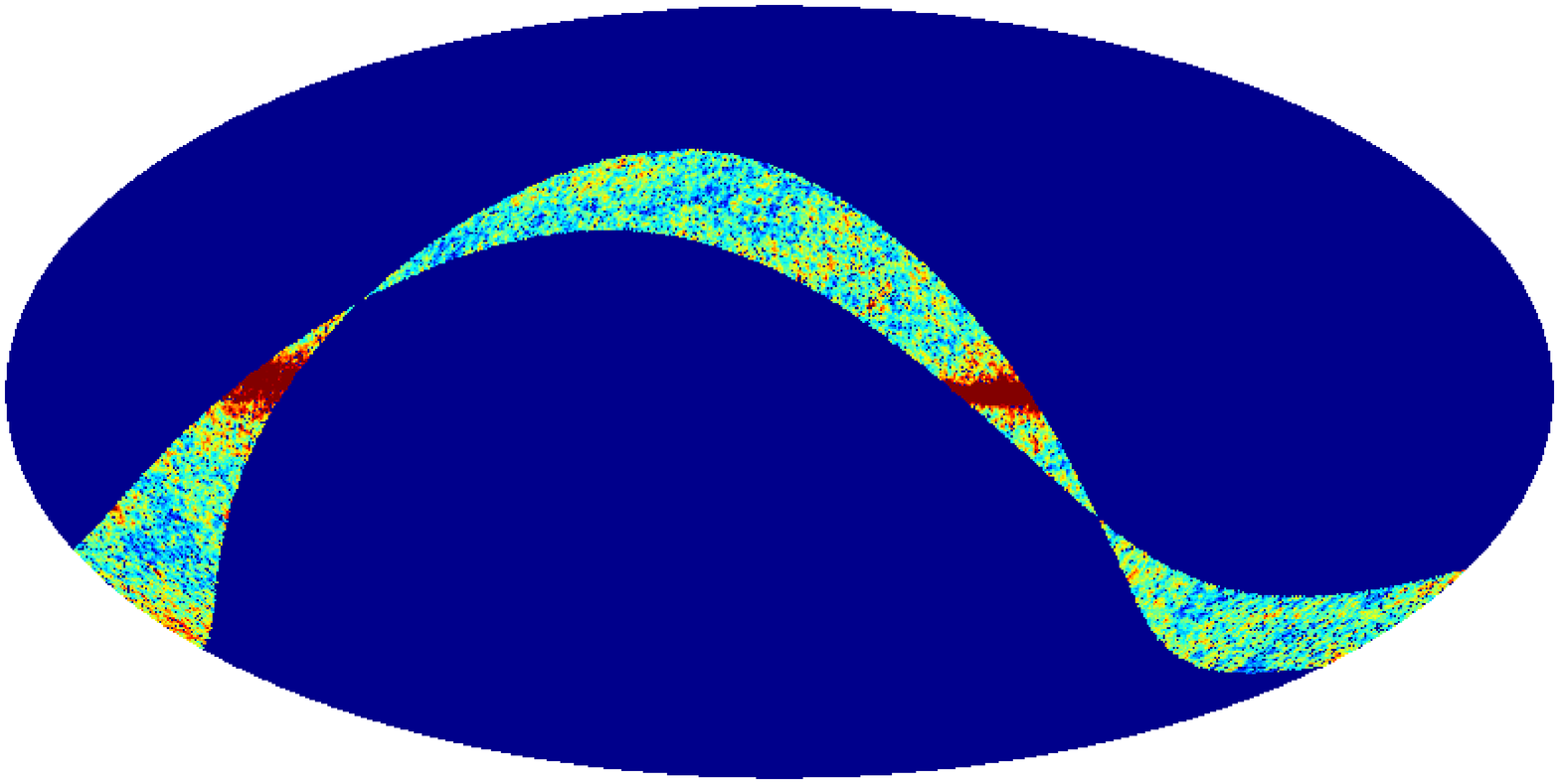}
\end{center}
\vspace{-4mm} \caption{The original Planck first light survey (upper panel)
and derived CMB temperature map (lower panel).}
\label{fig:image vs map}
\end{figure}

The reliability of our method can be checked by comparing
the Mollweide view image of the recovered map with the initial Planck image.
As shown in Fig.~\ref{fig:image vs map}, they are consistent to each other.
We also test the relative error induced by the transmission from a JPEG 
image into fits format temperature map in this way: use the WMAP 5-year ILC map 
to plot a JPEG format Mollweide image; transform it into a $N_{side}=256$ 
fits format output temperature map with the same process used for the Planck 
first light image; compute the CMB power spectra of both the input ILC map 
and output map; then we find that the RMS of relative difference between 
the two power spectra is about $15\%$ for $l=2-350$.

\subsection{From color bin number to temperature}
\label{sub:From color bin number to real temperatures}
To transform the color bin number to real temperatures,
three methods are adopted to estimate the temperature to bin number ratio $R$.

Firstly, we know that for pure CMB, the RMS is about 50\,$\mu$K,
and the Planck first light survey covers about $10\%$ of the sky 
or more than 70,000 effective pixels;
therefore, the maximum and minimum temperatures in the image should be
at least $\pm3\sigma$, which is about $\pm 150$ $\rm{\mu K}$ in 256 bins.
This gives the lower limit of the ratio $R$ which is 1.17 $\rm{\mu K}/bin$.
Since there exist foreground and noise, the real ratio must be
significantly higher than this value.

Secondly, for all pixels in the Planck scanned region and outside
the WMAP KQ85 mask$^{[5]}$, we sort their color bin numbers and take
out the maximum $10\%$ and minimum $10\%$ to compute the average values,
which are 58.7 and 172.9 respectively. Then we do the same thing
to the 5-year WMAP ILC map and get -123 and 118\,$\rm{\mu K}$. Therefore,
the ratio $R$ estimated in this way is roughly 2.11\,$\rm{\mu K}/bin$.
With other WMAP bands from Q to W, $R$ varies from 2.1 to 2.9\,$\rm{\mu K}/bin$.

\begin{figure}[t]
\begin{center}
\includegraphics[width=0.45\textwidth,angle=270]{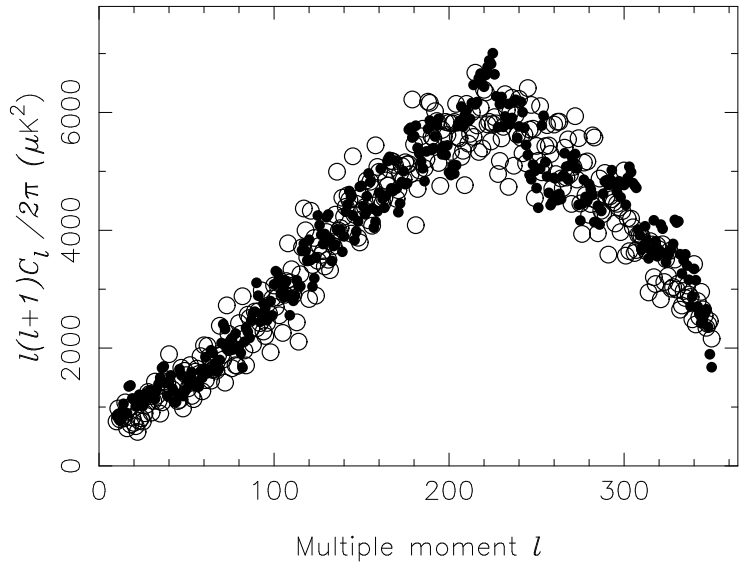}
\end{center}
\vspace{-4mm} \caption{Power spectra. {\sl Circle}: the released WMAP5 power spectrum. 
{\sl Filled circle}: the "bin number power spectrum" 
with the best fit parameters to match the WMAP5 power spectrum. 
}
\label{fig:bin}
\end{figure}

The last method is to compute the "bin number power spectrum",
which is to use the color bin number map instead of the temperatures
to compute the CMB power spectrum.
Then we fit the obtained power spectrum to known WMAP5 CMB power spectrum 
in range $l=10-350$ to determine the ratio $R$. Two models are used in fitting: 
The first is the WMAP5 CMB power spectrum, and the second is a white noise component. 
Both models are modified by a beam profile of 0-30 arcminute FWHM. 
The best least-square fit occurs at beam width equal to 22 arcminute FWHM 
and $R=2.96$ \,$\rm{\mu K}/bin$, which is very close to the value 
$R=2.87$ \,$\rm{\mu K}/bin$ derived by method 2 from W-band. 
The "bin number power spectrum" modified with the best-fit parameters 
is given in Fig.~\ref{fig:bin}. The maximum ratio is $R=3.3$ \,$\rm{\mu K}/bin$ 
and occurs at 0 arcminute FWHM, which can be taken as the upper limit of $R$.

In summary, the ratio $R$ is estimated with three independent methods and 
we get consistent results including the best-fit value and 
the upper and lower limits. This is useful, because we find that, 
even with the upper limit of $R$, the Planck low-$l$ components 
(proportional to $R^2$) are still significantly lower than WMAP. 
Therefore we can safely conclude that the Planck low-$l$ 
components are really lower than WMAP.

\section{CMB power spectrum}
\label{sec:The CMB power spectrum from "Planck"}

With the "Planck" CMB temperature map obtained above, we start to compute
the CMB power spectrum. We use a mask to remove the unscanned pixels
(deep blue pixels in the lower panel of Fig.~\ref{fig:image vs map})
and then use the WMAP 5-year KQ85 mask$^{[5]}$ to remove the pixels
possibly contaminated by strong foreground emission.
After that, a CMB power spectrum is derived from the map. 
This CMB power spectrum still need to be fixed for beam function 
and noises; however, there is no such information and the only way (perhaps) 
is to do some fittings like in method 3 in \S\ref{sub:From color bin number to real temperatures}. 
The result is actually the same as shown in Fig.~\ref{fig:bin}. 
We don't provide more power spectrum for $l>350$ because the effect 
of beam function and noises are strong there, and we need the true 
Planck beam function and noise properties to obtain reliable results 
for such high-$l$ components.

Not like the high-$l$ components, the low-$l$ components are almost unaffected 
by the beam function or noises, and we have seen substantial differences 
between the Planck and WMAP published maps
at these low-$l$ (large angular scale) components.

Since the Planck scanned region is only about $10\%$ of the sky, 
it is difficult to compare its low-$l$ components directly with
the released WMAP results;
however, we can use the \textbf{same mask} (the Planck unscanned region + KQ85)
to compute the power spectra from the WMAP (V-band) and our safe-mode (also V-band) maps,
and completely neglect the cut-sky correction.
In this way, low-$l$ components in the scanned region
can be directly compared under exactly the same conditions.
The results are listed in Table~1.
The ratio $R$ used in Table~1 is the upper limit: 3.3 $\rm{\mu K}/bin$. 
Note that the power spectrum is proportional to $R^2$. 
The uncertainties are estimated by the results from three different bands of WMAP. 
We deduce that the Planck uncertainties should have similar or most probably 
smaller amplitude.
From Table~1 we see that the ``Planck'' power spectrum at low-$l$ is apparently much lower 
than WMAP values even with the upper limit of $R$.
Compared to the WMAP official maps, our safe mode maps$^{[3]}$
are apparently more consistent with ``Planck'' at low-$l$.

\begin{table}{Table 1: Low-$l$ components of power spectra from CMB maps 
in the Planck first light survey region without cut-sky correction, in $\rm{\mu K}^2$.}\\[1ex]
  \label{tab:low-$l$ components}
  \begin{tabular}{ccccc}\hline
$l$ & Planck  &  WMAP (release) & WMAP (safe mode)  \\
\hline
2 & 1.91 & $2.56\pm 0.13$ & $1.91\pm 0.15$   \\
3 & 3.75 & $9.83\pm 0.47$ & $8.88\pm 0.41$   \\
4 & 3.94 & $6.25\pm 0.76$ & $4.35\pm 0.68$    \\
5 & 29.6 & $50.0\pm 0.65$ & $46.2\pm 0.55$    \\ \hline
\end{tabular}
\end{table}

Two points have to be noticed: firstly, the ratio $R$ is the upper limit, 
3.3 $\rm{\mu K}/bin$, in Table 1. This strongly indicates that the real 
Planck low-$l$ components should be even lower than listed. 
Secondly, it seems that the foreground emission has not been removed 
from the Planck first light survey image. By removing the foreground, 
it is certain that the Planck low-$l$ components will further decrease. 
For instance, the WMAP values in Table 1 is derived from foreground-clean map, 
otherwise the quadrupole will be 15.1 $\rm{\mu K}^2$, not merely 2.56 $\rm{\mu K}^2$.

\section{Effect of the systematics}
\label{sec:Effect of the systematics}

It is almost certain that the Planck first light survey image
contain some systematics. Here, the effect of different kinds 
of systematics upon the low-$l$ components are discussed.

Firstly, the beam profile and noise properties are unknown for the
first light survey image; however, the low-$l$ components are
almost unaffected by them. The ratio $R$ is affected; 
however, as discussed above, even with the upper limit of $R$, 
the low-$l$ components of Planck are still much lower than WMAP, 
to say nothing of the best-fit value of $R$.

Secondly, it seems that the foreground emission has not been removed
in the Planck image. However, after applying the KQ85 mask, the
residual foreground is actually very weak and does not affect the
power spectrum significantly, e.g., no more than $10\%$ according to
experience from WMAP, and most importantly, removing the foreground
will weaken both the high-$l$ and low-$l$ components, which will
only enhance our conclusion.

Next, the temperature map resolution is limited by the image
resolution and can not exceed $N_{side}=256$; however, this is quite
enough to estimate the low-$l$ components. In fact, $N_{side}=8$ is
already enough for $l\le 5$.

In the end, although the Planck image must have been well-calibrated
(as emphasized by them in bold in~[1], the first light
survey image is used to "demonstrate the ability to calibrate the
equipment over long periods"), it is still possible that the
calibration is imperfect. Although we don't know the details of
Planck calibration, it is reasonable to believe that, like WMAP, the
most important purpose of calibration is to determine the equipment
gain \footnote{No equipment can work like a thermometer and measure
the CMB temperature directly, so something like a gain between the CMB
radiation intensity and the temperature fluctuation must exist in all
CMB experiments} and baseline. The gain artifact affects both
high-$l$ and low-$l$ components equally. Since the Planck
high-$l$ components are very close to expectation (Fig.~\ref{fig:bin}), we can safely
conclude that the gain artifacts must be weak. The baseline
artifact is very similar to the foreground, and we can even call it
"equipment foreground"; therefore, same as the foreground,
cleaning the gain artifact will only make the low-$l$ components
lower and our conclusion stronger.

It is common to see in experiments that, lower is the
systematics, lower is the derived values, especially for something
like energy or power spectrum. That's because errors or systematics
are expected to contribute an extra "power", which must be uncorrelated with the true
signal, and the overall effect will be most probably "larger". If in
the Planck first light survey, it is the systematics that
make the low-$l$ components too lower, then the systematics must
be significantly correlated with the true signal, which is almost impossible. 
Therefore, we believe that the Planck
low-$l$ components are really lower than WMAP.

\section{Discussion}
\label{sec:conclusion}
From the Planck first light sky survey image we see
that the most prominent feature of CMB anisotropy, revealed by WMAP
with its first acoustic peak, is certainly real
and well confirmed by Planck. On the other hand, 
for the low-$l$ components (see Table~1), 
significant differences between the Planck preliminary spectrum
and WMAP release exist. It is obvious that the published Planck first light image 
is possibly contaminated by systematic instrumental effects and 
might be unsuitable for precise cosmology study, e.g.
for cosmological parameters estimation. 
However, as discussed in \S~\ref{sec:Effect of the systematics}, 
for the low-$l$ components, the fact that they are significantly 
lower than WMAP is quite beyond doubt.

Recently, we found a remarkable inconsistency between
the input data and reconstructed temperatures in WMAP team's map-making
and observed serious ecliptic contamination in published temperature maps$^{[3]}$.
That indicates there must be a serious problem in the map making routine of the WMAP team.
We then independently developed a map-making software which can
completely overcome the problem of inconsistency.
We used the improved software to produce new WMAP temperature
maps in safe mode with stronger constraints on data selection for planet
avoidance. The power spectrum derived from our safe mode maps
has nearly zero quadrupole component, which is consistent with
what revealed by the Planck first light survey.

Tegmark et al.$^{[6]}$ found in the released first year WMAP maps
both the CMB quadrupole ($l=2$) and octopole ($l=3$)
having power along a particular spatial axis, and more works$^{[7-10]}$
found that the preferred directions
of these two low-$l$ components are highly aligned and close to the
ecliptic plane. Such a strange orientation of large-scale patterns of CMB maps in
respect to the ecliptic frame has long puzzled cosmologists.
Our work$^{[3]}$ pointed out that the CMB quadrupole component apparent in
the published WMAP maps is not of cosmological origin, 
but mainly came from  ecliptic contamination
induced by inappropriate map-making. From the comparison between Planck and WMAP
for greyscale maps by Cover$^{[2]}$
and for power spectra in our present work, the Planck first light survey image
strongly supports the artificial origin of quadrupole observed in official WMAP
maps and the real CMB quadrupole is most possibly near zero.

Table~1 shows that not only the quadrupole, but also other low-$l$ components,
from $l=3$ (octopole) to $l=5$, are very weak.
We have found two kinds of systematic errors in official WMAP maps.
One is the pixels in the WMAP scan ring of a hot pixel being
systematically cooled$^{[11,12]}$.  Another one is the
significant correlation between the temperature
and the observation number of a sky pixel$^{[11,13]}$.
These two kinds of systematic effects also remain in our safe mode maps.
The sky coverage of WMAP mission is nonuniform: the number of observations
is greatest near the ecliptic poles and in rings $\sim 141\degree$ from each pole,
and least in the ecliptic plane.
The non-uniform exposure pattern should produce artificial power component
at low-$l$ through the systematic effects mentioned above.
It is expected that after properly correcting systematics
 low-$l$ spectrum of WMAP safe-mode maps can be
more consistent with the Planck survey result.


\end{document}